\documentclass[fleqn,twoside]{article}
\usepackage{espcrc2}

\usepackage{graphicx}
\usepackage[figuresright]{rotating}

\newcommand{\AmS}{{\protect\the\textfont2
  A\kern-.1667em\lower.5ex\hbox{M}\kern-.125emS}}

\title{
\vskip -110pt
{\large  
\mbox{} \hfill BNL-NT-03/19\\
\mbox{} \hfill BI-TP 2003/22\\
\mbox{} \hfill DESY 03-119\\
}
\vskip 45pt
Charmonium at finite temperature
       }

\author{P. Petreczky
\address[BNL]{Physics Department, Brookhaven National Laboratory,
        Upton, NY 11973 USA}
        \thanks{Goldhaber Fellow},                             
        S. Datta\address[BI]{
        Fakult\"at f\"ur Physik, Universit\"at Bielefeld,
        D-33615 Bielefeld, Germany}, F. Karsch\addressmark[BI],
        I. Wetzorke
        \address{
        NIC/DESY Zeuthen, Platanenallee 6, D-15738 Zeuthen,
        Germany 
        }
        }
       
\begin{document}

\begin{abstract}
We study charmoinum correlators and spectral functions 
at finite temperature within the quenched approximation using 
isotropic lattices with lattice spacing
$a^{-1}=4.86$~GeV and $9.72$~GeV. 
Although we observe some medium modifications
of the ground state charmonium spectral function above 
deconfinement, 
we find that ground state charmonia
($J/\psi$ and $\eta_c$) exist in the deconfined phase
at least up to temperatures as high as $1.5T_c$.  
$P$-wave charmonia ($\chi_c$) on the other hand are dissociated 
already at $1.12T_c$. 
\vspace{1pc}
\end{abstract}

\maketitle

\section{Introduction}
The study of charmonium at finite temperature
has received considerable interest ever since the 
seminal paper by Matusi and Satz \cite{MS86}. The 
behavior of charmoinum at finite temperature
was studied using potential models with some screened 
potential \cite{MS86,other} and it was found that
excited states like $\psi'$ and $\chi_c$ are dissociated
at $T_c$ while the ground state charmonia $J/\psi$ and
$\eta_c$ were found to be dissociated at $1.1T_c$.
In last year's conference (Lattice 2002)
lattice calculation of the charmonium
spectral functions were reported \cite{datta02,umeda02}
which indicated that in contrast to potential model
expectations ground state charmonia may exist up
to temperatures as high as $1.5T_c$ while $\chi_c$
states are dissociated \cite{datta02}. 
In this paper
we are going to present our results on the charmonium
system at finite temperature on isotropic lattice with
lattice spacing $a^{-1}=4.86$ and $9.72$ GeV. The 
spectral functions were extracted using the Maximum
Entropy Method (MEM) \cite{mem} and confirm earlier results on 
survival of the ground state and dissolution of 
$\chi_c$ states. The survival of the ground state 
charmonia has been confirmed also very recently in \cite{asakawa03}. 
For other studies of meson spectral
functions at finite temperature using MEM see
\cite{dil,asakawa02}.

\section{Numerical results}
We consider the temporal correlators of meson currents
\begin{equation}
G(\tau,T)=\sum_{\vec{x}} \langle J(\vec{x},\tau) J(\vec{0},0) \rangle~,
\end{equation}
with $J=\bar q(\vec{x},\tau) \Gamma q(\vec{x},\tau)$ and
$\Gamma=1,\gamma_5,\gamma_{\mu}, \gamma_5 \gamma_{\mu}$ for
scalar, pseudoscalar, vector and axial vector channels.
The low-lying states in these channels correspond to
$\chi_{c0}$, $\eta_c$, $J/\psi$ and $\chi_{c1}$ mesons,
respectively. The spectral function which we are interested in
is related to the meson correlator by the integral relation
\begin{equation}
G(\tau,T)=\int_0^{\infty} d \omega \sigma(\omega,T) 
\frac{\cosh \omega(\tau-1/(2T))}{\sinh \omega/(2T)}~.
\label{rel}
\end{equation}
Although this relation is valid a priori only in the continuum,
it has been shown recently that also in the limit of non-interacting 
theory it holds on the lattice \cite{karsch03}. In our
simulations we  used non-perturbatively improved
Wilson fermions \cite{luescher} and the standard Wilson action for
gauge fields. The simulation parameters are summarized in Table 1.
The $\kappa$ values in this table correspond to a vector meson mass
close to the actual $J/\psi$ mass, {\it i.e.}
$3.13(1)$~GeV at  $\beta=6.64$ and $3.83(2)$~GeV at $\beta=7.192$
(these were estimated from the spatial correlators). 
\begin{table}[htb]
\caption{ 
Simulation parameters.
$T_c$ denotes the deconfinement temperature.
}
\label{table:1}
\newcommand{\m}{\hphantom{$-$}}
\newcommand{\cc}[1]{\multicolumn{1}{c}{#1}}
\renewcommand{\arraystretch}{1.2} 
\begin{tabular}{@{}lllll}
\hline
 $\beta$  & \cc{$a^{-1}$ [GeV]} & \cc{$\kappa$} & \cc{$Size$}  & \cc{$T/T_c$} \\
\hline
6.640     & 4.86          &  0.1290       & $48^3 \times 24$ & 0.75          \\
          &               &               & $48^3 \times 16$ & 1.12          \\
          &               &               & $48^3 \times 12$ & 1.50          \\
7.192     & 9.72          &  0.13114      & $40^3 \times 40$ & 0.90          \\
          &               &               & $64^3 \times 24$ & 1.50          \\
          &               &               & $48^3 \times 12$ & 3.00          \\
\hline
\end{tabular}\\[2pt]
\vspace*{-0.9cm}
\end{table}
In Fig. 1 we show charmonium spectral functions below the deconfinement
temperature, $T_c$, reconstructed using MEM for two different lattice
spacings. One can clearly identify the ground state peak corresponding
to $J/\psi$, $\eta_c$ and $\chi_c$ states. The other structures in the
spectral functions are lattice artifacts as their position strongly 
depends on the lattice spacing (they scale roughly as $1/a$).
In our analysis we have used free lattice spectral functions 
from \cite{karsch03} as default model (for other forms, 
see Ref. \cite{mem}) in MEM analysis.

\begin{figure}[htb]
\includegraphics[width=6.6cm]{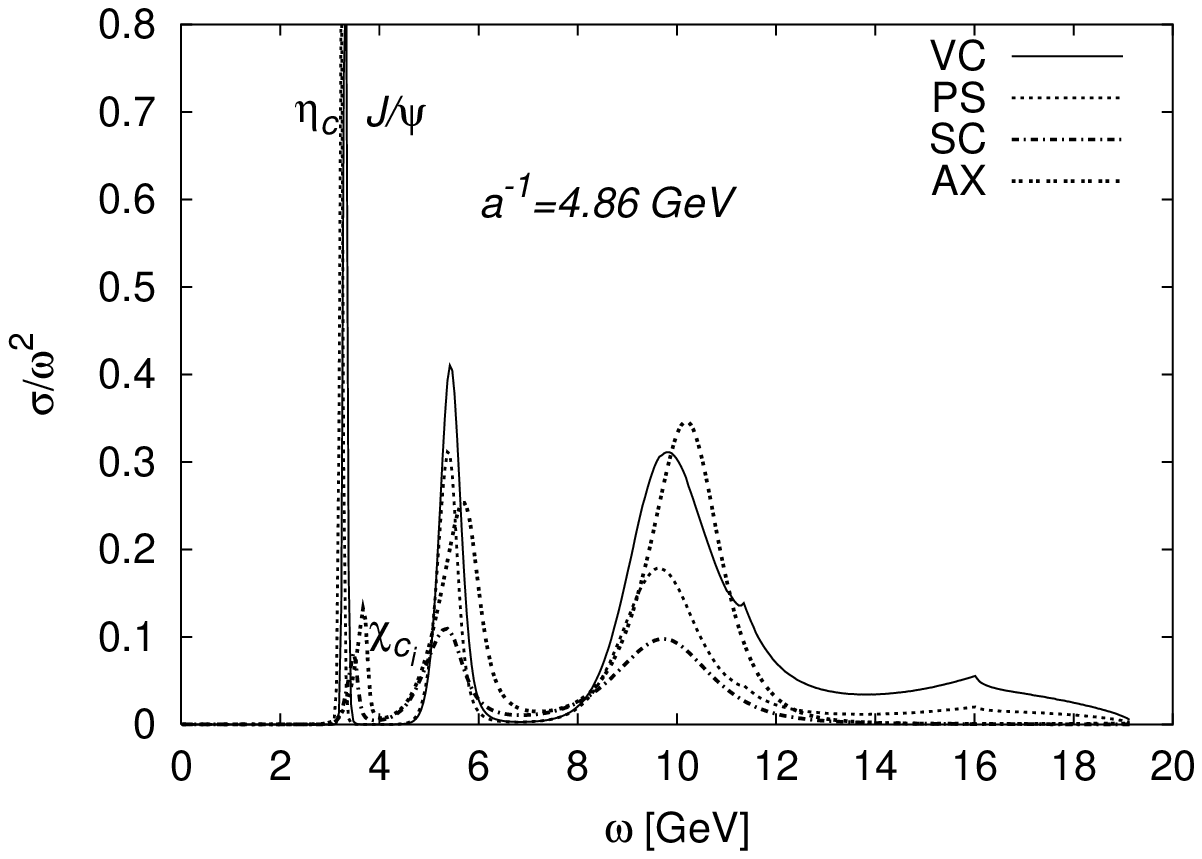}\\
\includegraphics[width=6.6cm]{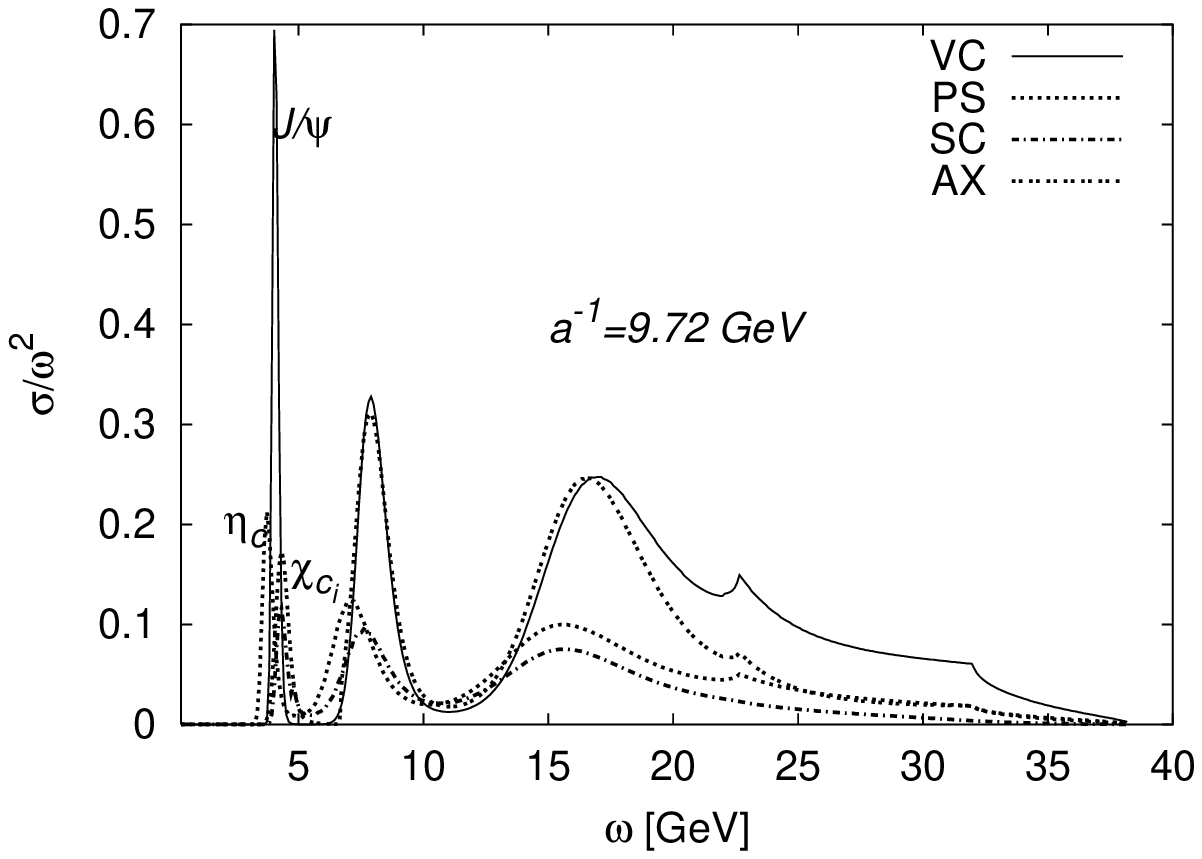}
\vspace*{-0.8cm}
\caption{Spectral functions at $0.75T_c$ and $a^{-1}=4.86$~GeV
(top) and $0.9T_c$ and $a^{-1}=9.72$~GeV (bottom).
}
\label{fig1}
\vspace*{-0.9cm}
\end{figure}
Below the deconfinement temperature the temporal extent of the lattice
is about 1 fm or larger and spectral functions can be
reliably reconstructed. As we increase the temperature
the temporal extent becomes smaller and 
smaller and the reconstruction of the spectral function
becomes more and more difficult.
However, even without reconstructing the spectral
function certain statements about their temperature dependence can be
made just by inspecting the temperature dependence of the
corresponding meson correlators. The temperature dependence
of the meson correlators comes from the temperature
dependence of the integration kernel in Eq. \ref{rel} and from
the temperature dependence of the spectral function. In order
to separate the former effect from the temperature dependence arising
from modifications of 
the spectral function we consider the ratios 
$G(\tau,T)/G_{recon}(\tau)$ for different channels with
$G_{recon}=\int d \omega \sigma(\omega,T^{*})\cosh (\omega(\tau-1/2T))/
\sinh(\omega/2T)$ and $T^{*}$ being the smallest temperature 
available at that
lattice spacing, i.e. $T^{*}=0.75T_c$ for the data set
with $a^{-1}=4.86$~GeV and $T^{*}=0.9T_c$ for data with  $a^{-1}=9.72$~GeV.
Our numerical results for $G/G_{recon}$ are shown
in Fig. 2.
\begin{figure}[htb]
\includegraphics[width=6.6cm]{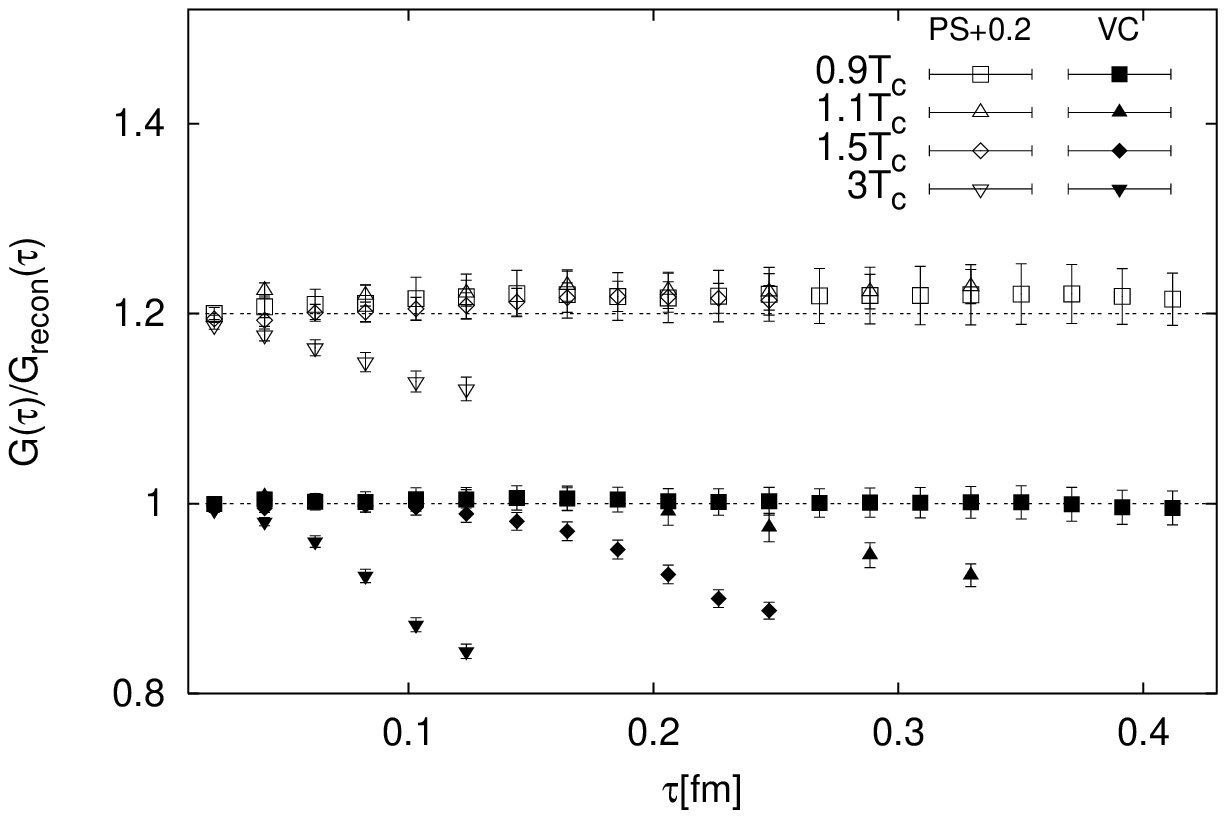}\\
\includegraphics[width=6.6cm]{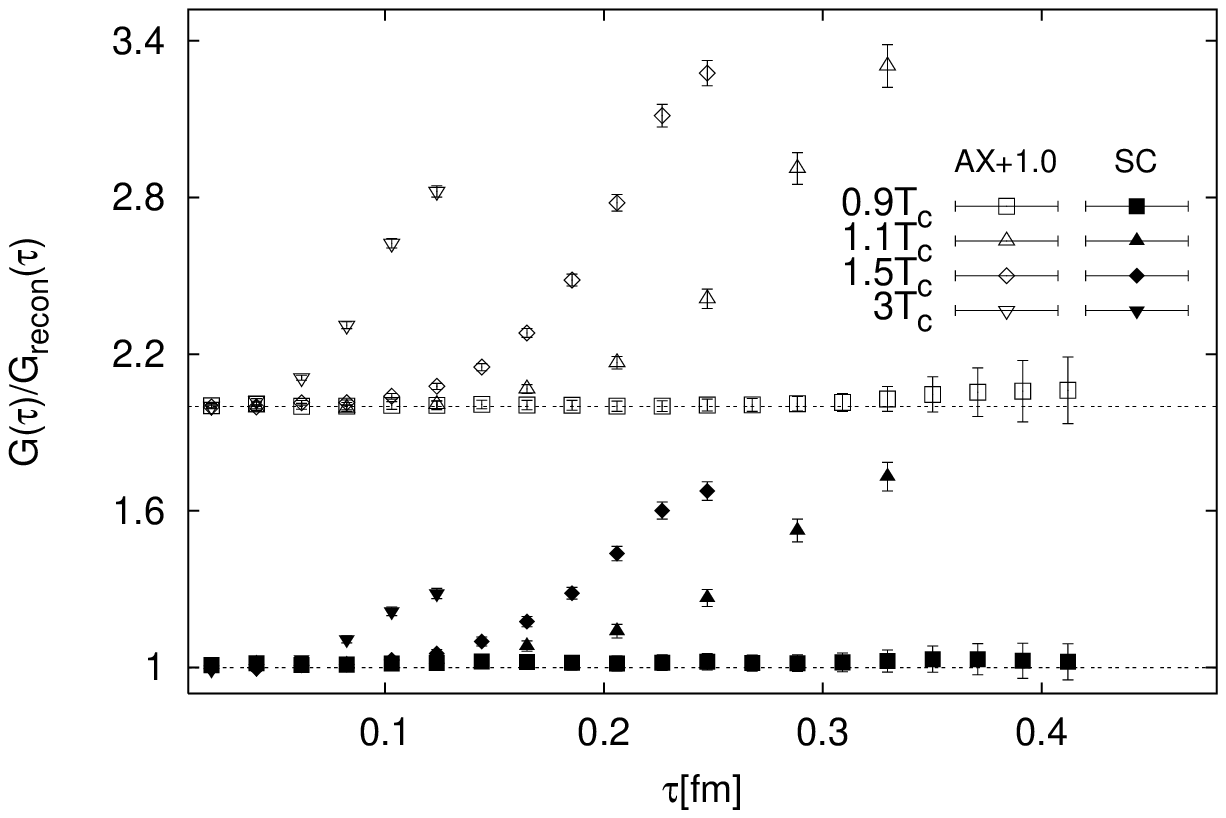}
\vspace*{-1.1cm}
\caption{$G/G_{recon}$ for different temperatures for
vector and pseudoscalar channels 
(top) and in the scalar and axial vector channels (bottom).
The data for the pseudoscalar and axial vector channels
have been shifted by a constant for better visualization.
}
\label{fig2}
\vspace*{-0.9cm}
\end{figure}
As one can see from the figure in the pseudoscalar channel
almost no modification occurs at the level of the correlation
functions up to temperatures $1.5T_c$ but substantial modifications
of the correlator are seen at $3T_c$.
In the vector channel
we see small gradual modifications of the correlators with increasing
temperature which become substantial at $3T_c$. 
In contrast to this a strong temperature dependence is seen 
in the scalar and axial vector channels
as we cross $T_c$.

Now let us discuss the spectral functions above $T_c$. Since 
the physical extent of the imaginary time direction is rather small
at high temperature a reliable reconstruction of the spectral
function is difficult. In particular one finds that the 
reconstructed spectral functions are quite sensitive to the
default model. Thus a correct choice of the default model is
important. As we do not expect medium effects at energies 
considerably larger than the open charm threshold we 
will use the high energy part of the 
spectral function calculated below $T_c$ as a default model. At low
energy this is matched to a  default model of the form 
$m(\omega)=m_1 \omega^2$,
with $m_1$ chosen such that $m(\omega)$ is a continuous
function in the entire energy interval. The vector and scalar
spectral functions
above $T_c$ are shown in Fig. 3. For comparison
we also plot there the spectral function at $0.9T_c$.
As one can see the ground state peak in the vector channel
(the $J/\psi$ state) exists in the deconfined phase up to
the temperature $T=1.5T_c$ while there is no such peak in
the scalar channel even at $1.1T_c$, i.e. the $\chi_{c0}$ state
is dissolved already at that temperature. The calculation of
the errors shown in Fig. 3 is explained in \cite{mem}.

\begin{figure}[htb]
\includegraphics[width=6.6cm]{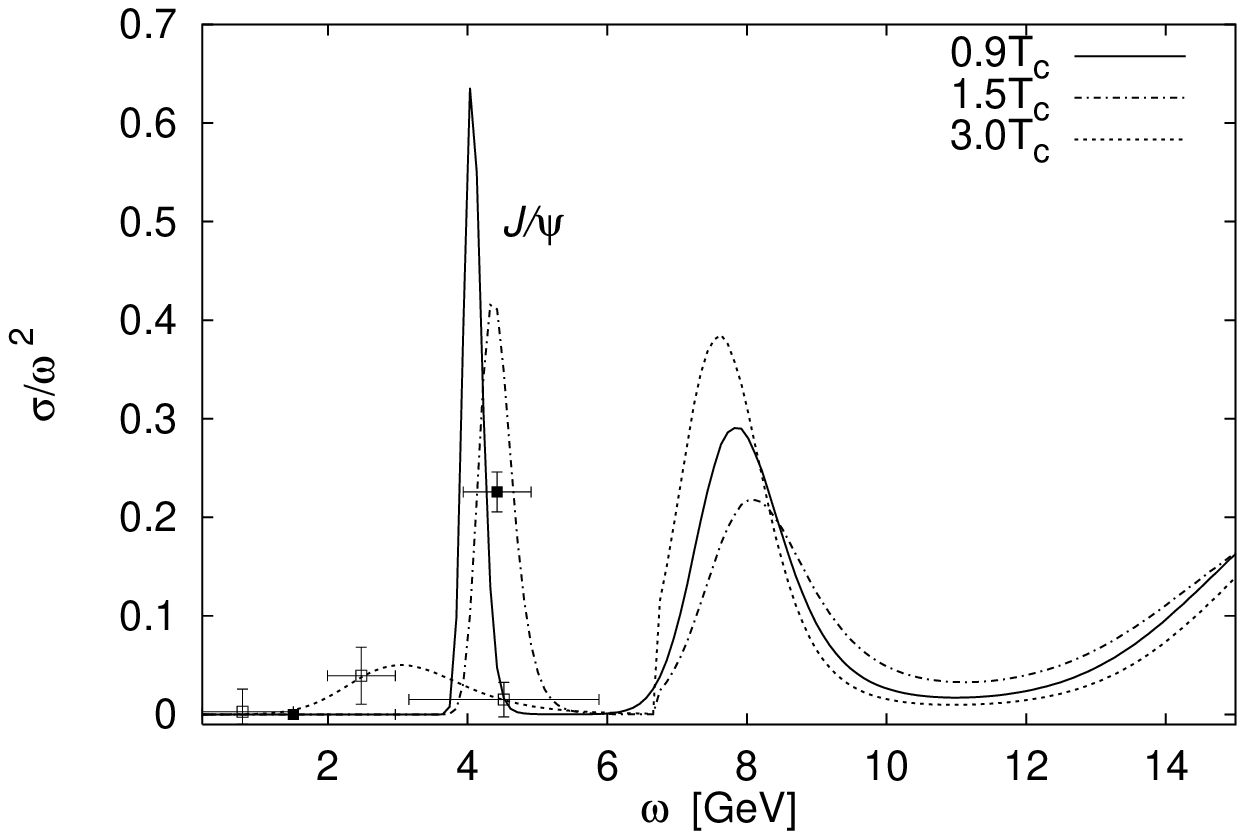}\\
\includegraphics[width=6.6cm]{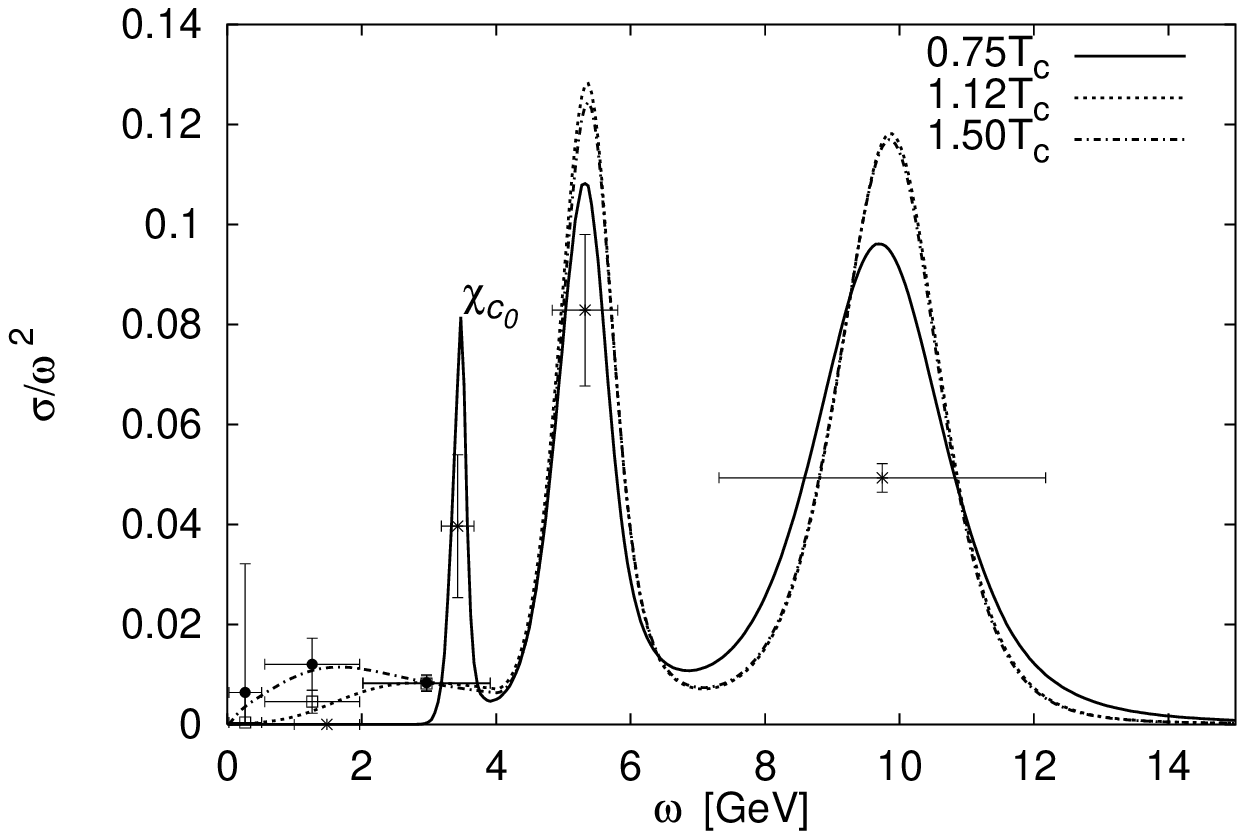}
\vspace*{-0.9cm}
\caption{Spectral functions above and below
deconfinement in the vector at $\beta=7.192$ (top) 
and scalar channel at $\beta=6.64$ (bottom). 
}
\label{fig3}
\end{figure}

\section*{Acknowledgment}
\noindent
This work has been authored under the contract 
DE-AC02-98CH10886 with the U.S. Department of energy.
Support from DFG under grant FOR 339/2-1 is gratefully
acknowledged.


\begin{thebibliography}{9}
\bibitem{MS86} T. Matsui and H. Satz, Phys. Lett. {\bf B178} (1986) 416
\bibitem{other} F. Karsch et al, Z. Phys. {\bf C37} (1988) 617;
S. Digal et al, Phys. Rev. D {\bf 64} (2001) 094015 
\bibitem{datta02} S. Datta et al, hep-lat/0208012
\bibitem{umeda02} T. Umeda et al, hep-lat/0209139; see also 
hep-lat/0211003 for a more detailed analysis 
\bibitem{mem}
Y. Nakahara et al, Phys. Rev. D {\bf 60} (1999) 091503;
M. Asakawa et al, Prog. Part. Nucl. Phys. {\bf 46} (2001) 459
\bibitem{asakawa03}
M. Asakawa and T. Hatsuda, hep-lat/0308034
\bibitem{dil} 
F. Karsch et al, Phys. Lett. {\bf B530} (2002) 147
\bibitem{asakawa02}
M. Asakawa et al, hep-lat/0209059
\bibitem{karsch03}
F. Karsch et al, Phys. Rev. D {\bf 68} (2003) 014504
\bibitem{luescher}
M. L\"uscher et al, Nucl. Phys. {\bf B469} (1996) 419
\end{thebibliography}
\end{document}